\documentstyle[12pt,a4,psfig]{article}
\begin{document}
\title{The entropy multiparticle-correlation expansion \\
for a mixture of spherical and elongated particles}
\author{\\
Santi Prestipino~\cite{aff} and Paolo V. Giaquinta \\ \\
{\it Universit\`a degli Studi di Messina} \\
{\it and} \\
{\it Istituto Nazionale per la Fisica della Materia (INFM)} \\ \\
{\it Dipartimento di Fisica, Contrada Papardo, 98166 Messina, Italy}
}
\date{}
\maketitle
\begin{abstract}
We derive the multiparticle-correlation expansion of the excess
entropy of classical particles in the canonical ensemble using a
new approach that elucidates the rationale behind each term in the
expansion. This formula provides the theoretical framework for an
entropy-based ordering criterion that has been already tested for
a variety of model fluids and thermodynamic phenomena. In view of
future investigations of the phase diagram of colloidal mixtures,
we detail in this paper the case of a two-component system of
spherical and rod-like particles and discuss the symmetries
underlying both the self and distinct pair-distribution functions
under various geometrical constraints.
\end{abstract}

\vspace{5mm}

\vspace{5mm}

\vspace{5mm}
\noindent Running title: {\em Entropy multiparticle-correlation expansion
for binary mixtures}
\thispagestyle{empty}

\newpage
%
%
\section{Introduction}
\setcounter{page}{1}

One-phase criteria have been often introduced to estimate the
location of the phase boundaries of the liquid and solid
phases~\cite{criteria}. These empirical rules are usually rather
specific. Nevertheless, they can be quite useful when it is not
easy or straightforward to evaluate the free energies of the
competing phases. Among such rules, one formerly proposed by
Giaquinta and coworkers~\cite{Giaquinta} has proved to be a rather
general and flexible tool that can be successfully applied to a
variety of models under different structural and thermodynamic
conditions. This rule actually qualifies as an ordering criterion
that can be readily implemented on the basis of the properties of
the more disordered phase. The theoretical framework is provided
by the multiparticle-correlation expansion (MPCE) of the excess
entropy of a classical fluid that was originally derived by H.~S.
Green in the canonical ensemble~\cite{Green}, and was later
extended by Nettleton and M.~S. Green to an open
system~\cite{Nettleton}. According to this formula, the entropy
can be written as an infinite series whose $n$-th term is the
contribution associated with density correlations involving
$n$-particle multiplets. In short, the criterion states that the
overall contribution to the entropy of a fluid arising from the
spatial correlations involving {\it more than two} particles -- a
quantity called ``residual multi-particle entropy'' (RMPE) --
changes sign concurrently with the local emergence of any new kind
of structural organization in the system. The zero-RMPE criterion
has been tested against such diverse thermodynamic phenomena as
freezing~\cite{Giaquinta,freezing1,freezing2}, fluid-fluid phase
separation in hard-sphere mixtures~\cite{Saija}, mesophase
formation in model liquid crystals~\cite{Costa,Cuetos}, the
Kosterlitz and Thouless metal-insulator transition in a
two-dimensional Coulomb lattice gas~\cite{Donato}, and, more
recently, the density-maximum anomaly in liquid
water~\cite{Saitta}.

In a previous paper, we gave an entirely new proof of the entropy
MPCE in the canonical ensemble by exploiting a simple
combinatorial identity~\cite{Prestipino}. The proof applies to
both continuous and lattice systems. In this paper we reconsider
this derivation from a different perspective which discloses the
significance of each term in the expansion. Moreover, we outline
an iterative method for building up the expansion term by term.
This method is then extended to systems composed of two different
species of particles. Indeed, the present physical motivation for
developing this formalism, besides the self-standing interest in
the formal derivation of an entropy MPCE for binary mixtures, is
that of demonstrating the effectiveness and reliability of the
zero-RMPE criterion also in the case of a colloidal mixture of
spherical and rod-like particles. In particular, we plan to check
whether the miscible, low-density phase of such a model undergoes
a kind of structural instability -- of the type indicated by the
vanishing of the RMPE -- towards a more ordered phase, be it {\em
lamellar} (for low concentrations of the spheres) or {\em
immiscible} (the bulk-phase-separated system), for values of the
total packing fraction close to those independently ascertained by
experiment and numerical simulation~\cite{Adams,Dogic}. Previous
studies have already shown that the RMPE of parallel
spherocylinders vanishes at about the same density where the
smectic order sets in~\cite{Costa}. We are confident that the
effects on the phase behavior that are associated with the
addition of a small quantity of spheres will be correctly
accounted for by the RMPE.

While deferring to a forthcoming paper the discussion of the
numerical simulation study of a mixture of hard spheres and
spherocylinders, we anticipate here the analysis of the symmetries
owned by the three pair distribution functions that enter the
calculation of the two-body entropy.

%
%
\section{One-component systems}
\renewcommand{\theequation}{2.\arabic{equation}}

We start discussing the entropy MPCE of one-component systems in
the canonical ensemble. This choice is by no means restrictive
since, as Baranyai and Evans first pointed out, it is always
possible to take advantage of the canonical sum rules of the
correlation functions and then rearrange the entropy MPCE in an
ensemble-invariant form~\cite{Baranyai}.

Let ${\bf R}^N=\{{\bf R}_1,\ldots,{\bf R}_N\}$ and ${\bf
P}^N=\{{\bf P}_1,\ldots,{\bf P}_N\}$ be the set of all particle
coordinates and momenta, respectively. The canonical partition
function of the system can always be split into an ideal and an
excess part, $Z_N=Z_N^{\rm id}Z_N^{\rm ex}$, where:
\begin{equation}
Z_N^{\rm id}=\frac{1}{N!}\left( \frac{V}{\Lambda^3}\right) ^N
\,\,\,\,\,\,{\rm and}\,\,\,\,\,\,
Z_N^{\rm ex}=\frac{1}{V^N}
\int{\rm d}^3R_1\ldots{\rm d}^3R_N\,e^{-\beta V_N({\bf R}^N)}\,.
\label{2-01}
\end{equation}
In Eq.\,(\ref{2-01}), $\beta=1/(k_BT)$, $\Lambda=h/\sqrt{2\pi mk_BT}$,
$V$ is the volume, and $V_N({\bf R}^N)$ is an arbitrary potential-energy
function (in the most general case, $V_N$ is a sum of $n$-body
terms with $n=1,2,\ldots,N$).
The excess entropy $S_N^{\rm ex}\equiv S_N-S_N^{\rm id}$ reads:
\begin{equation}
\frac{S_N^{\rm ex}}{k_B}=
-\int\frac{{\rm d}^3R_1\ldots{\rm d}^3R_N}{V^N}\,
\frac{e^{-\beta V_N({\bf R}^N)}}{Z_N^{\rm ex}}
\ln\frac{e^{-\beta V_N({\bf R}^N)}}{Z_N^{\rm ex}}\,.
\label{2-02}
\end{equation}
Upon defining a set of $N$ normalized distribution functions (DFs) as:
\begin{eqnarray}
P^{(N)}({\bf R}^N) &=& \frac{e^{-\beta V_N({\bf R}^N)}}{Z_N^{\rm ex}}\,;
\nonumber \\
P^{(n)}({\bf R}^n) &=&
\int\frac{{\rm d}^3R_{n+1}\ldots{\rm d}^3R_N}{V^{N-n}}\,
\frac{e^{-\beta V_N({\bf R}^N)}}{Z_N^{\rm ex}}
\,\,\,\,\,\,\,\,\,\,(n=1,\ldots,N-1)\,,
\label{2-03}
\end{eqnarray}
with the properties
\begin{equation}
\int\frac{{\rm d}^3R_1\ldots{\rm d}^3R_n}{V^n}\,P^{(n)}({\bf R}^n)=1
\,\,\,\,\,\,{\rm and}\,\,\,\,\,\,
\int\frac{{\rm d}^3R_{n+1}}{V}\,P^{(n+1)}({\bf R}^{n+1})=
P^{(n)}({\bf R}^n)\,,
\label{2-04}
\end{equation}
the ordinary $n$-body DFs can be expressed as:
\begin{eqnarray}
f^{(n)}({\bf r}^n) &\equiv& \left< \sum_{i_1\ldots i_n}^{\prime}
\delta^3({\bf R}_{i_1}-{\bf r}_1)\ldots\delta^3({\bf R}_{i_n}-{\bf
r}_n)\right>
\nonumber \\
&=& \frac{N!}{(N-n)!}
\left< \delta^3({\bf R}_1-{\bf r}_1)\ldots\delta^3({\bf R}_n-{\bf r}_n)\right>
\nonumber \\
&=& \frac{N!}{(N-n)!}\,\frac{P^{(n)}({\bf r}^n)}{V^n}\,,
\label{2-05}
\end{eqnarray}
where the sum is carried out over all $n$-tuples of
distinct particle labels. The reduced $n$-body DFs (for
$n=2,\ldots,N$) read:
\begin{eqnarray}
g^{(n)}({\bf r}^n) &\equiv&
\frac{f^{(n)}({\bf r}^n)}{f^{(1)}({\bf r}_1)\cdots f^{(1)}({\bf r}_n)}
\nonumber \\
&=& \left( 1-\frac{1}{N}\right)
\cdots\left( 1-\frac{n-1}{N}\right) \frac{P^{(n)}({\bf r}^n)}
{P^{(1)}({\bf r}_1)\cdots P^{(1)}({\bf r}_n)}
\nonumber \\
&\equiv& \prod_{a=1}^n\left(1-\frac{a-1}{N}\right)
\tilde{P}^{(n)}({\bf r}^n)\,, \label{2-06}
\end{eqnarray}
and verify the property:
\begin{equation}
\int\frac{{\rm d}^3R_{n+1}}{V}\,P^{(1)}({\bf R}_{n+1})g^{(n+1)}({\bf R}^{n+1})=
\left( 1-\frac{n}{N}\right) g^{(n)}({\bf R}^n)\,,
\label{2-07}
\end{equation}
which holds also for $n=1$ if $g^{(1)}\equiv 1$. We note that
$P^{(1)}=1$ and $f^{(1)}=N/V$ if no one-body term is present in
$V_N$, {\it i.e.}, if no external potential acts on the particles
(for the ideal gas, $P^{(n)}=1$ for all $n$). From now on, we
shall adopt the shorthand notations $P_{12\ldots n}=P^{(n)}({\bf
R}^n)$ and $\tilde{P}_{12\ldots n}=\tilde{P}^{(n)}({\bf R}^n)$.
Moreover, any integral of the form
$V^{-n}\int{\rm d}^3R_1\ldots{\rm d}^3R_n\,(\cdots)$ is hereafter
denoted as $\int(\cdots)$.

Now, we show how to build up the MPCE of the entropy term by term.
The strategy is to consider a progressively increasing number of
particles in the system. For $N=1$, the (adimensional) excess
entropy is just $S_1^{\rm ex}/k_B=-\int P_1\ln P_1$, which leads
to a first-order approximation to the excess entropy of a
$N$-particle system, $S_N^{\rm ex}/k_B\approx S_N^{(1)}/k_B\equiv
-N\int P_1\ln P_1$ ({\it{viz.}}, each particle contributes to the
entropy independently of the other particles).

Next, we move to a system of two particles only, and write its
excess entropy as $S_2^{(1)}$ plus a remainder $k_BR_2$ that is
equal to:
\begin{equation}
R_2\equiv\frac{S_2^{\rm ex}-S_2^{(1)}}{k_B}=-\int P_{12}\ln P_{12}+
2\int P_1\ln P_1=-\int P_{12}\ln\tilde{P}_{12}\,.
\label{2-08}
\end{equation}
Equation (\ref{2-08}) leads to a second-order approximation for
$S_N^{\rm ex}$ under the hypothesis that each distinct pair of
particles contributes an equal two-body residual term to the
entropy:
\begin{equation}
\frac{S_N^{(2)}}{k_B}=-N\int P_1\ln P_1- {N\choose 2}\int
P_{12}\ln\tilde{P}_{12}\,. \label{2-09}
\end{equation}
Note that the approximation (\ref{2-09}) is exact for $N=2$, {\it
i.e.}, $S_2^{(2)}=S_2^{\rm ex}$.

When there are three particles in the system, the excess entropy
is the sum of $S_3^{(2)}$ and a remainder $k_BR_3$:
\begin{equation}
R_3\equiv\frac{S_3^{\rm ex}-S_3^{(2)}}{k_B}=
-\int P_{123}\ln\tilde{P}_{123}+{3\choose 2}\int P_{12}\ln\tilde{P}_{12}\,,
\label{2-10}
\end{equation}
which suggests a third-order approximation for $S_N^{\rm ex}$ in
the form
\begin{equation}
\frac{S_N^{(3)}}{k_B}=-N\int P_1\ln P_1-
{N\choose 2}\int P_{12}\ln\tilde{P}_{12}-
{N\choose 3}\left[ \int P_{123}\ln\tilde{P}_{123}-
{3\choose 2}\int P_{12}\ln\tilde{P}_{12}\right] \,.
\label{2-11}
\end{equation}
Again, note that $S_3^{(3)}=S_3^{\rm ex}$. Equation (\ref{2-11})
reproduces the first three terms in the r.h.s. of Eq.\,(5.9) of
Ref.\,\cite{Prestipino}, and one may suspect that the remaining
terms will be similarly discovered by arguing for $N=4,5,\ldots$
as we did for $N=1,2,3$. The related proof can be obtained by
induction over $N$. In fact, we know our target~\cite{Prestipino}:
\begin{eqnarray}
\frac{S_N^{\rm ex}}{k_B} &=& -N\int P_1\ln P_1-
\int P_{12\ldots N}\ln\tilde{P}_{12\ldots N}
\nonumber \\
&=& -N\int P_1\ln P_1-\sum_{n=2}^N{N\choose n}\sum_{a=2}^n(-1)^{n-a}
{n\choose a}\int P_{1\ldots a}\ln\tilde{P}_{1\ldots a}\,.
\label{2-12}
\end{eqnarray}
After taking $I_a\equiv\int P_{1\ldots a}\ln\tilde{P}_{1\ldots a}$
and $J_a\equiv\int P_{1\ldots a}\ln P_{1\ldots a}$ (for
$a=1,\ldots,N$), we first show that Eq.\,(\ref{2-12}) is correct
or, equivalently, that:
\begin{equation}
I_N=\sum_{n=2}^N{N\choose n}\sum_{a=2}^n(-1)^{n-a}{n\choose a}I_a\,.
\label{2-13}
\end{equation}
To this aim, it is enough to observe that the coefficient of $I_a$
in the sum (\ref{2-13}) is (cf. Eq.\,(5.10) of \cite{Prestipino}):
\begin{equation}
\sum_{n=a}^N(-1)^{n-a}{N\choose n}{n\choose a}=\left\{
\begin{array}{rl}
0\,, & \,\,\,{\rm for}\,\,0\le a<N \\
1\,, & \,\,\,{\rm for}\,\,a=N
\end{array}
\right.
\label{2-14}
\end{equation}

Equation (\ref{2-12}) can be built up term by term using the same
procedure as sketched above for $N=1,2,3$. Upon defining for a
given $M$ (with $2\le M\le N-1$):
\begin{equation}
\frac{S_N^{(M)}}{k_B}=-NJ_1-
\sum_{n=2}^M{N\choose n}\sum_{a=2}^n(-1)^{n-a}{n\choose a}I_a\,,
\label{2-15}
\end{equation}
which is tantamount to truncating the sum (\ref{2-12}) over $n$
after the $M$-th term, we set $S_{M+1}^{\rm ex}\equiv
S_{M+1}^{(M)}+k_BR_{M+1}$, where:
\begin{eqnarray}
R_{M+1} &=& -(\underbrace{J_{M+1}-(M+1)J_1}_{I_{M+1}})+
\sum_{n=2}^M{M+1\choose n}\sum_{a=2}^n(-1)^{n-a}{n\choose a}I_a
\nonumber \\
&=& -\sum_{n=2}^{M+1}{M+1\choose n}\sum_{a=2}^n(-1)^{n-a}{n\choose a}I_a+
\sum_{n=2}^M{M+1\choose n}\sum_{a=2}^n(-1)^{n-a}{n\choose a}I_a
\nonumber \\
&=& -\sum_{a=2}^{M+1}(-1)^{M+1-a}{M+1\choose a}I_a\,.
\label{2-16}
\end{eqnarray}
This result allows one to define a higher-order approximation to the
entropy as:
\begin{eqnarray}
\frac{S_N^{(M+1)}}{k_B} &\equiv& \frac{S_N^{(M)}}{k_B}+{N\choose M+1}R_{M+1}
\nonumber \\
&=& -NJ_1-\sum_{n=2}^M{N\choose n}\sum_{a=2}^n(-1)^{n-a}{n\choose a}I_a-
{N\choose M+1}\sum_{a=2}^{M+1}(-1)^{M+1-a}{M+1\choose a}I_a
\nonumber \\
&=& -NJ_1-\sum_{n=2}^{M+1}{N\choose n}\sum_{a=2}^n(-1)^{n-a}{n\choose a}I_a\,,
\label{2-17}
\end{eqnarray}
which is exactly the same as (\ref{2-15}), but for the quantity
$M+1$ which replaces $M$.

The MPCE of the entropy remains formally the same if the particles
possess further degrees of freedom besides those pertaining to
the centre of mass. For instance, in the case of liquid-crystal
molecules ({\it i.e.}, elongated particles with cylindrical
symmetry), there are two more degrees of freedom for each particle
since two angular coordinates are needed to specify the
orientation of the molecule in three-dimensional space. If the
body $z$ axis is taken to coincide with the molecular axis, we can
represent the direction of a molecule by resorting to the Euler
angles $\theta$ and $\phi$ (see Fig.\,1). The third angle, $\psi$,
describes a rotation around the molecular axis and, as such, it is
not relevant for the configuration of the molecule. Let $\xi=({\bf
R},\theta,\phi)$ be the five-dimensional vector of coordinates of
an individual molecule. The interaction potential is then a
function of $\xi^N$.

One of the simplest reference models for a liquid crystal is a
system of hard spherocylinders or, equivalently, rods that cannot
approach each other beyond a given distance $\sigma$ (the
spherocylinder diameter). Such particles show up-down symmetry.
In this specific case, the potential $V_N(\xi^N)$ will also reflect
this symmetry, in that it must be invariant upon interchanging
$(\theta,\phi)$ with $(\pi-\theta,\pi+\phi)$.

The rotational kinetic energy $K_{\rm rot}$ of a massive rod,
written in canonical coordinates, reads:
\begin{equation}
K_{\rm rot}=\frac{P_\theta^2}{2I}+\frac{P_\phi^2}{2I\sin^2\theta}\,,
\label{2-18}
\end{equation}
where $I$ is the moment of inertia relative to any axis
perpendicular to the rod and passing through its centre. It then
easily follows that the ideal and excess partition functions are:
\begin{equation}
Z_N^{\rm id}=\frac{1}{N!}\left( \frac{4\pi V}{\lambda^2\Lambda^3}\right) ^N
\,\,\,\,\,\,{\rm and}\,\,\,\,\,\,
Z_N^{\rm ex}=\frac{1}{(4\pi V)^N}\int{\rm d}^5\xi_1\ldots{\rm d}^5\xi_N\,
e^{-\beta V_N(\xi^N)}\,,
\label{2-19}
\end{equation}
where $\lambda=h/\sqrt{2\pi Ik_BT}$ and ${\rm
d}^5\xi=\sin\theta\,{\rm d}^3R\,{\rm d}\theta\,{\rm d}\phi$. The
factor $\sin\theta$ in the volume element originates from the
Gaussian integral over $P_\phi$. Because of this factor, the delta
function of argument $\xi$ should be intended as follows:
\begin{equation}
\delta^5(\xi-\xi_1)\equiv\delta^3({\bf R}-{\bf R}_1)\,
\frac{\delta(\theta-\theta_1)}{\sin\theta}\,\delta(\phi-\phi_1)\,.
\label{2-20}
\end{equation}
With this proviso, the formal definition of the DFs, given by
Eq.\,(\ref{2-05}), as well as the overall appearance of the entropy
formula (see Eq.\,(\ref{2-12})), remain unchanged provided that we
now interpret $\int(\cdots)$ as
$(4\pi V)^{-n}\int{\rm d}^5\xi_1\ldots{\rm d}^5\xi_n\,(\cdots)$
(the value of $n$ is always implicit in the form of the integrand).

Let us now consider the properties of the reduced pair
distribution function (PDF) in relation to the symmetries of the
system. Assume that no external field is present and that the
molecular interaction includes, besides the hard-core repulsion,
at most a pair term whose strength depends on the distance
between the centres of mass.
In such a case, the PDF $g^{(2)}(\xi_1,\xi_2)$ will only depend on
the relative position of the two molecules. In a reference frame
$\Sigma_1$ where molecule 1 is placed at the origin and oriented
along the $z$ axis, the position of molecule 2 is thoroughly
described by the three (spherical) coordinates of its centre of
mass ($r_{12},\vartheta_{12},$ and $\varphi_{12}$) plus two more
angles, $\theta_{12}$ and $\phi_{12}$, specifying the direction
of its axis. However, the above description is redundant since the
orientation of, say, the $x$ axis of $\Sigma_1$ is still arbitrary
and we can always arrange things in such a way that
$\varphi_{12}=0$. Hence, $g^{(2)}$ will depend on four variables
only, namely one distance ($r_{12}$) and three angles
($\vartheta_{12},\theta_{12},$ and $\phi_{12}$).

An even simpler situation is that of an artificially constrained
nematic fluid, namely a system of elongated particles whose axes
are kept parallel to each other while their centres of mass are
free to move. In this case, the angles $\theta_{12}$ and
$\phi_{12}$ are no longer necessary, with the result that the PDF
depends just on $r_{12}$ and $\vartheta_{12}$ (or, equivalently,
on $\rho_{12}=r_{12}\sin\vartheta_{12}$ and
$z_{12}=r_{12}\cos\vartheta_{12}$).

Another interesting case is that of an inhomogeneous system of
rod-like molecules confined by an impenetrable wall. This model is
useful for investigating the wetting properties of a nematic fluid
and the onset and growth of a smectic layer at the wall. If the
strength of the wall-particle attraction depends just on the
distance $z$ of the molecular centre of mass from the wall, the
number density $f^{(1)}(\xi)$ turns out to be a function of
$\chi=(z,\theta,\phi)$ ({\it i.e.}, the axis direction is relevant
even when the wall does not exert any attraction), while the PDF
$g^{(2)}(\xi_1,\xi_2)$ depends on $\chi_1$ and $\chi_2$.

%
%
\section{Two-component systems}
\renewcommand{\theequation}{3.\arabic{equation}}
\setcounter{equation}{0}

In this section, we generalize the MPCE of the entropy to binary
systems composed of two different kinds of particles. We shall
proceed in two steps: we shall first use the iterative method
outlined in the previous section as a guidance for conjecturing
the complete formula from its first few terms. Then, we shall give
a formal proof of this formula by induction over the total number
of particles.

The canonical partition function of a two-component system with
$N=N_1+N_2$ classical point particles generally reads
$Z_{N_1,N_2}=Z_{N_1}^{\rm id,\,1}Z_{N_2}^{\rm id,\,2}Z_{N_1,N_2}^{\rm ex}$,
where the excess part has the form:
\begin{equation}
Z_{N_1,N_2}^{\rm ex}=
\frac{1}{V^N}\int{\rm d}^{3N_1}R\,{\rm d}^{3N_2}Q\,
e^{-\beta V_{N_1,N_2}({\bf R}^{N_1},\,{\bf Q}^{N_2})}\,.
\label{3-01}
\end{equation}
In Eq.\,(\ref{3-01}), the potential-energy function is arbitrary.
The excess entropy is given by the integral
\begin{equation}
\frac{S_{N_1,N_2}^{\rm ex}}{k_B}=
-\frac{1}{V^N}\int{\rm d}^{3N_1}R\,{\rm d}^{3N_2}Q\,%
\frac{e^{-\beta V_{N_1,N_2}({\bf R}^{N_1},\,{\bf Q}^{N_2})}}
{Z_{N_1,N_2}^{\rm ex}}%
\ln\frac{e^{-\beta V_{N_1,N_2}({\bf R}^{N_1},\,{\bf Q}^{N_2})}}
{Z_{N_1,N_2}^{\rm ex}}\,.
\label{3-02}
\end{equation}
Upon defining
\begin{eqnarray}
P^{(N_1,N_2)}({\bf R}^{N_1},{\bf Q}^{N_2}) &=&
\frac{e^{-\beta V_{N_1,N_2}({\bf R}^{N_1},\,{\bf Q}^{N_2})}}
{Z_{N_1,N_2}^{\rm ex}}\,;
\nonumber \\
P^{(n_1,\,n_2)}({\bf R}^{n_1},{\bf Q}^{n_2}) &=&
\int\frac{{\rm d}^3R_{n_1+1}\ldots{\rm d}^3R_{N_1}}{V^{N_1-n_1}}\,%
\frac{{\rm d}^3Q_{n_2+1}\ldots{\rm d}^3Q_{N_2}}{V^{N_2-n_2}}
\frac{e^{-\beta V_{N_1,N_2}({\bf R}^{N_1},\,{\bf Q}^{N_2})}}
{Z_{N_1,N_2}^{\rm ex}}
\nonumber \\
&& ({\rm with}\,\,n_1\le N_1,\,n_2\le N_2,\,n_1+n_2=1,\ldots,N-1)\,,
\label{3-03}
\end{eqnarray}
we have the following properties:
\begin{eqnarray}
&& \int\frac{{\rm d}^3R_1\ldots{\rm d}^3R_{n_1}}{V^{n_1}}\,%
\frac{{\rm d}^3Q_1\ldots{\rm d}^3Q_{n_2}}{V^{n_2}}\,%
P^{(n_1,\,n_2)}({\bf R}^{n_1},{\bf Q}^{n_2})=1\,;
\nonumber \\
&& \int\frac{{\rm d}^3R_{n_1+1}}{V}\,%
P^{(n_1+1,\,n_2)}({\bf R}^{n_1+1},{\bf Q}^{n_2})=
P^{(n_1,\,n_2)}({\bf R}^{n_1},{\bf Q}^{n_2})\,;
\nonumber \\
&& \int\frac{{\rm d}^3Q_{n_2+1}}{V}\,%
P^{(n_1,\,n_2+1)}({\bf R}^{n_1},{\bf Q}^{n_2+1})=
P^{(n_1,\,n_2)}({\bf R}^{n_1},{\bf Q}^{n_2})\,.
\label{3-04}
\end{eqnarray}
When external fields are absent, $P^{(1,\,0)}=1$ and
$P^{(0,\,1)}=1$. For a binary ideal mixture ({\it i.e.},
$V_{N_1,N_2}=0$), the $P$ functions are all equal to 1.

The DF of order $(n_1,\,n_2)$ is defined as:
\begin{equation}
f^{(n_1,\,n_2)}({\bf r}^{n_1},{\bf q}^{n_2})=
\frac{N_1!}{(N_1-n_1)!}\,\frac{N_2!}{(N_2-n_2)!}\,%
\frac{P^{(n_1,\,n_2)}({\bf r}^{n_1},{\bf q}^{n_2})}{V^{n_1+n_2}}\,.
\label{3-05}
\end{equation}
While the definition of the {\em self} reduced DFs $g^{(n_1,\,0)}$
(for $n_1\ge 1$) and $g^{(0,\,n_2)}$ (for $n_2\ge 1$) is strictly
analogous to that given for a one-component system, the {\em
distinct} reduced DFs are defined, for $n_1,n_2\ge 1$, as:
\begin{equation}
g^{(n_1,\,n_2)}({\bf r}^{n_1},{\bf q}^{n_2})=
\prod_{a=1}^{n_1}\left( 1-\frac{a-1}{N_1}\right)
\prod_{b=1}^{n_2}\left( 1-\frac{b-1}{N_2}\right)
\tilde{P}^{(n_1,\,n_2)}({\bf r}^{n_1},{\bf q}^{n_2})\,,
\label{3-06}
\end{equation}
where
\begin{equation}
\tilde{P}^{(n_1,\,n_2)}({\bf r}^{n_1},{\bf q}^{n_2})\equiv
\frac{P^{(n_1,\,n_2)}({\bf r}^{n_1},{\bf q}^{n_2})}
{P^{(1,\,0)}({\bf r}_1)\cdots P^{(1,\,0)}({\bf r}_{n_1})
P^{(0,\,1)}({\bf q}_1)\cdots P^{(0,\,1)}({\bf q}_{n_2})}\,.
\label{3-07}
\end{equation}
For $n_1+n_2\ge 1$, a property analogous to that expressed in
Eq.\,(\ref{2-07}) holds:
\begin{equation}
\int\frac{{\rm d}^3r_{n_1+1}}{V}\,P^{(1,\,0)}({\bf r}_{n_1+1})
g^{(n_1+1,\,n_2)}({\bf r}^{n_1+1},{\bf q}^{n_2})=
\left( 1-\frac{n_1}{N_1}\right) g^{(n_1,\,n_2)}({\bf r}^{n_1},{\bf q}^{n_2})\,,
\label{3-08}
\end{equation}
plus a similar identity involving $g^{(n_1,\,n_2+1)}$.

>From now on, a short notation is adopted where $P_{ab}$ stands for
$P^{(a,\,b)}({\bf r}^a,{\bf q}^b)$ and any integral of the form
$V^{-n_1-n_2}\int{\rm d}^{3n_1}r\,{\rm d}^{3n_2}q\,(\cdots)$ is
simply denoted as $\int(\cdots)$. Moreover, for future
convenience, we take $I_{ab}\equiv\int P_{ab}\ln\tilde{P}_{ab}$
and $J_{ab}\equiv\int P_{ab}\ln P_{ab}$ (with $a\le N_1,\,b\le
N_2,\,a+b\ge 1$). For instance, the excess entropy (\ref{3-02})
can be rewritten as
\begin{eqnarray}
\frac{S_{N_1,N_2}^{\rm ex}}{k_B} &=& -\int P_{N_1N_2}\ln P_{N_1N_2}
\nonumber \\
&=& -N_1\int P_{10}\ln P_{10}-N_2\int P_{01}\ln P_{01}-
\int P_{N_1N_2}\ln\tilde{P}_{N_1N_2}\,,
\label{3-09}
\end{eqnarray}
or $J_{N_1N_2}=N_1J_{10}+N_2J_{01}+I_{N_1N_2}$.

We now move on to determine a MPCE for the entropy of a mixture.
To this aim, we consider increasing particle numbers in the
system, starting from $N_1+N_2=1$ ({\it i.e.}, only one particle,
either of type 1 or 2, is present). By reasoning in the usual way,
we immediately obtain a first-order approximation to the excess
entropy in the form
\begin{equation}
\frac{S_{N_1,N_2}^{(1)}}{k_B}=
-N_1\int P_{10}\ln P_{10}-N_2\int P_{01}\ln P_{01}\,.
\label{3-10}
\end{equation}

Then, we analyze the three cases with $N_1+N_2=2$. For $N_1=2$ and
$N_2=0$ (or the other way round), things are the same as for a
one-component system, {\it i.e.}, $S_{2,\,0}^{\rm ex}=
S_{2,\,0}^{(1)}+k_BR_{2,0}$, with $R_{2,0}=-\int
P_{20}\ln\tilde{P}_{20}$. Instead, for $N_1=N_2=1$:
\begin{equation}
R_{1,1}\equiv\frac{S_{1,\,1}^{\rm ex}-S_{1,\,1}^{(1)}}{k_B}=
-\int P_{11}\ln P_{11}+\int P_{10}\ln P_{10}+\int P_{01}\ln P_{01}=
-\int P_{11}\ln\tilde{P}_{11}\,,
\label{3-11}
\end{equation}
thus leading to the following second-order approximation for
$S_{N_1,N_2}^{\rm ex}$:
\begin{eqnarray}
\frac{S_{N_1,N_2}^{(2)}}{k_B} &=&
-N_1\int P_{10}\ln P_{10}-N_2\int P_{01}\ln P_{01}
\nonumber \\
&-& {N_1\choose 2}\int P_{20}\ln\tilde{P}_{20}-
N_1N_2\int P_{11}\ln\tilde{P}_{11}-
{N_2\choose 2}\int P_{02}\ln\tilde{P}_{02}\,.
\nonumber \\
\label{3-12}
\end{eqnarray}

For $N_1+N_2=3$, we just reproduce below the expression of
$S_{N_1,N_2}^{(3)}$, which results from carefully considering the
implied four cases: \footnotesize
\begin{eqnarray}
&& \frac{S_{N_1,N_2}^{(3)}}{k_B}=
-N_1\int P_{10}\ln P_{10}-N_2\int P_{01}\ln P_{01}
\nonumber \\
&& -{N_1\choose 2}\int P_{20}\ln\tilde{P}_{20}-
N_1N_2\int P_{11}\ln\tilde{P}_{11}-
{N_2\choose 2}\int P_{02}\ln\tilde{P}_{02}
\nonumber \\
&& -{N_1\choose 3}\left[ \int P_{30}\ln\tilde{P}_{30}-
{3\choose 2}\int P_{20}\ln\tilde{P}_{20}\right]
-{N_1\choose 2}N_2\left[ \int P_{21}\ln\tilde{P}_{21}-
\int P_{20}\ln\tilde{P}_{20}-2\int P_{11}\ln\tilde{P}_{11}\right]
\nonumber \\
&& -N_1{N_2\choose 2}\left[ \int P_{12}\ln\tilde{P}_{12}-
2\int P_{11}\ln\tilde{P}_{11}-\int P_{02}\ln\tilde{P}_{02}\right]
-{N_2\choose 3}\left[ \int P_{03}\ln\tilde{P}_{03}-
{3\choose 2}\int P_{02}\ln\tilde{P}_{02}\right]
\nonumber \\
\label{3-13}
\end{eqnarray}
\normalsize

On the basis of the above structure, we conjecture the following
general formula:
\begin{eqnarray}
\frac{S_{N_1,N_2}^{\rm ex}}{k_B} &=&
-N_1\int P_{10}\ln P_{10}-N_2\int P_{01}\ln P_{01}
\nonumber \\
&-& \sum_{\shortstack{\scriptsize $n_1+n_2\ge 2$ \\
\scriptsize $n_1\le N_1,\,n_2\le N_2$}}
{N_1\choose n_1}{N_2\choose n_2}
\sum_{\shortstack{\scriptsize $a+b\ge 2$ \\
\scriptsize $a\le n_1,\,b\le n_2$}}
(-1)^{n_1+n_2-a-b}{n_1\choose a}{n_2\choose b}\int P_{ab}\ln\tilde{P}_{ab}\,,
\nonumber \\
\label{3-14}
\end{eqnarray}
which was preliminary checked against the form of
$S_{N_1,N_2}^{(4)}$ as independently got by our iterative method.
Note that the double sum in Eq.\,(\ref{3-14}) can also be arranged
in such a way that all the terms with the same value of $n\equiv
n_1+n_2$ are gathered together:
\begin{equation}
I_{N_1N_2}=\sum_{n=2}^{N_1+N_2}\sum_{n_1=\max\{n-N_2,\,0\}}^{\min\{n,\,N_1\}}
{N_1\choose n_1}{N_2\choose n-n_1}
\sum_{\shortstack{\scriptsize $a+b\ge 2$ \\
\scriptsize $a\le n_1,\,b\le n-n_1$}}
(-1)^{n-a-b}{n_1\choose a}{n-n_1\choose b}I_{ab}\,.
\label{3-15}
\end{equation}

First of all, we prove that the Eq.\,(\ref{3-14}) is an exact identity.
Indeed, for given $a$ and $b$ (with $a\le N_1,\,b\le N_2,\,\,a+b\ge 2$),
the coefficient of $I_{ab}$ on the r.h.s. of (\ref{3-14}) is:
\begin{eqnarray}
&& -\sum_{\shortstack{\scriptsize $n_1+n_2\ge 2$ \\
\scriptsize $a\le n_1\le N_1,\,b\le n_2\le N_2$}}
{N_1\choose n_1}{N_2\choose n_2}\times
(-1)^{n_1+n_2-a-b}{n_1\choose a}{n_2\choose b}
\nonumber \\
&=& -\sum_{n_1=a}^{N_1}(-1)^{n_1-a}{N_1\choose n_1}{n_1\choose a}\times
\sum_{n_2=b}^{N_2}(-1)^{n_2-b}{N_2\choose n_2}{n_2\choose b}\,.
\label{3-16}
\end{eqnarray}
On account of Eq.\,(\ref{2-14}), the number (\ref{3-16}) is always
zero but for the case $(a,b)=(N_1,N_2)$, where it equals $-1$.
Hence, Eq.\,(\ref{3-14}) is correct.

We next prove, by induction, that Eq.\,(\ref{3-14}) also follows
from our iterative method. We define for any $M$ such that $2\le
M\le N-1$:
\begin{eqnarray}
\frac{S_{N_1,N_2}^{(M)}}{k_B} &=& -N_1J_{10}-N_2J_{01}
\nonumber \\
&-& \sum_{n=2}^M\,\sum_{n_1=\max\{n-N_2,\,0\}}^{\min\{n,\,N_1\}}
{N_1\choose n_1}{N_2\choose n-n_1}
\sum_{\shortstack{\scriptsize $a+b\ge 2$ \\
\scriptsize $a\le n_1,\,b\le n-n_1$}}
(-1)^{n-a-b}{n_1\choose a}{n-n_1\choose b}I_{ab}\,.
\nonumber \\
\label{3-17}
\end{eqnarray}
Moreover, for each pair $(M_1,M_2)$ satisfying $M_1+M_2=M+1$, let us put
$S_{M_1,M_2}^{\rm ex}=S_{M_1,M_2}^{(M)}+k_BR_{M_1,M_2}$, where:
\begin{eqnarray}
&& R_{M_1,M_2}=-(\underbrace{J_{M_1M_2}-M_1J_{10}-M_2J_{01}}_{I_{M_1M_2}})
\nonumber \\
&+& \sum_{n=2}^M\,\sum_{n_1=\max\{n-M_2,\,0\}}^{\min\{n,\,M_1\}}
{M_1\choose n_1}{M_2\choose n-n_1}
\sum_{\shortstack{\scriptsize $a+b\ge 2$ \\
\scriptsize $a\le n_1,\,b\le n-n_1$}}
(-1)^{n-a-b}{n_1\choose a}{n-n_1\choose b}I_{ab}
\nonumber \\
&=& -\sum_{n=2}^{M+1}\,\sum_{n_1=\max\{n-M_2,\,0\}}^{\min\{n,\,M_1\}}
{M_1\choose n_1}{M_2\choose n-n_1}
\sum_{\shortstack{\scriptsize $a+b\ge 2$ \\
\scriptsize $a\le n_1,\,b\le n-n_1$}}
(-1)^{n-a-b}{n_1\choose a}{n-n_1\choose b}I_{ab}
\nonumber \\
&+& \sum_{n=2}^M\,\sum_{n_1=\max\{n-M_2,\,0\}}^{\min\{n,\,M_1\}}
{M_1\choose n_1}{M_2\choose n-n_1}
\sum_{\shortstack{\scriptsize $a+b\ge 2$ \\
\scriptsize $a\le n_1,\,b\le n-n_1$}}
(-1)^{n-a-b}{n_1\choose a}{n-n_1\choose b}I_{ab}
\nonumber \\
&=& -\sum_{n_1=M+1-M_2}^{M_1}{M_1\choose n_1}{M_2\choose M+1-n_1}
\sum_{\shortstack{\scriptsize $a+b\ge 2$ \\
\scriptsize $a\le n_1,\,b\le M+1-n_1$}}
(-1)^{M+1-a-b}{n_1\choose a}{M+1-n_1\choose b}I_{ab}
\nonumber \\
&=& -\sum_{\shortstack{\scriptsize $a+b\ge 2$ \\
\scriptsize $a\le M_1,\,b\le M_2$}} (-1)^{M+1-a-b}{M_1\choose
a}{M_2\choose b}I_{ab}\,. \label{3-18}
\end{eqnarray}
In the third line of the above equation, the identity (\ref{3-15})
was used for $I_{M_1M_2}$. In the end, the same structure of
$S_{N_1,N_2}^{(M)}$ emerges for $S_{N_1,N_2}^{(M+1)}$ ({\it i.e.},
Eq.\,(3.17) where $M+1$ replaces $M$), as long as we define, in
close analogy with the one-component case:
\begin{equation}
\frac{S_{N_1,N_2}^{(M+1)}}{k_B}=\frac{S_{N_1,N_2}^{(M)}}{k_B}+
\sum_{M_1=\max\{M+1-N_2,\,0\}}^{\min\{M+1,\,N_1\}}
{N_1\choose M_1}{N_2\choose M+1-M_1}R_{M_1,M+1-M_1}\,.
\label{3-19}
\end{equation}
This concludes our proof.

Equations (\ref{2-12}) and (\ref{3-14}) express the canonical MPCE
of the excess entropy in compact form for pure and mixed systems,
respectively. In the literature, the expressions quoted for the
first few terms of this expansion are far more involved because
they are usually written in terms of the {\em reduced} DFs. We
have already shown in \cite{Prestipino} that the familiar form of
the entropy formula for one-component systems in the canonical
ensemble emerges when the $P$'s are eliminated in favour of the
$g$'s through the reverse of Eq.\,(\ref{2-06}). In doing so, a
constant term can be extracted from the correlation integrals
which, if absorbed into the ideal-gas part of the entropy, makes
the latter equivalent to the entropy of the {\em infinite-sized}
ideal-gas system (for the sake of clarity, this argument is
reformulated in appendix A).

Before providing a similar demonstration for mixed systems,
let us see what happens to $S_{N_1,N_2}^{(2)}$.
Upon repeatedly using Eqs.\,(\ref{3-04}) and (\ref{3-08}), we obtain
(with $\rho_1=N_1/V$ and $\rho_2=N_2/V$):
\begin{eqnarray}
&& \frac{S_{N_1,N_2}^{(2)}}{k_B}=-\rho_1\int{\rm d}^3r\,P_{10}({\bf r})\ln
P_{10}({\bf r})-\rho_2\int{\rm d}^3r\,P_{01}({\bf r})\ln P_{01}({\bf r})
\nonumber \\
&-& {N_1\choose 2}\ln\frac{N_1}{N_1-1}-\frac{1}{2}\rho_1^2
\int{\rm d}^3r\,{\rm d}^3r'\,P_{10}({\bf r})P_{10}({\bf r}')
g_{20}({\bf r},{\bf r}')\ln g_{20}({\bf r},{\bf r}')
\nonumber \\
&-& \rho_1\rho_2\int{\rm d}^3r\,{\rm d}^3r'\,P_{10}({\bf r})P_{01}({\bf r}')
g_{11}({\bf r},{\bf r}')\ln g_{11}({\bf r},{\bf r}')
\nonumber \\
&-& {N_2\choose 2}\ln\frac{N_2}{N_2-1}-\frac{1}{2}\rho_2^2
\int{\rm d}^3r\,{\rm d}^3r'\,P_{01}({\bf r})P_{01}({\bf r}')
g_{02}({\bf r},{\bf r}')\ln g_{02}({\bf r},{\bf r}')\,,
\nonumber \\
\label{3-20}
\end{eqnarray}
where all terms, including the two constants, are {\em extensive},
{\it viz.}, each term scales in the thermodynamic limit linearly
with either $N$ or $V$.

However, in a closed system the asymptotic value of the two-body
{\em self} reduced DFs differs at large distances from 1 for
${\cal O}(N^{-1})$ terms~\cite{Lebowitz}. This makes the numerical
evaluation of the integrals in (\ref{3-20}) particularly harmful
for a small system ({\it i.e.}, sensitive to its boundary). To
(partially) cure this problem, we can take advantage of the
canonical sum rules for the DFs (see Eqs.\,(\ref{3-08})), adding
(and subtracting) to each integral precisely the extensive term
that makes the integrand of order $N^{-2}$ at infinity:
\begin{eqnarray}
&& \frac{S_{N_1,N_2}^{(2)}}{k_B}=-\rho_1\int{\rm d}^3r\,P_{10}({\bf r})\ln
P_{10}({\bf r})-\rho_2\int{\rm d}^3r\,P_{01}({\bf r})\ln P_{01}({\bf r})
\nonumber \\
&-& \frac{1}{2}\rho_1^2
\int{\rm d}^3r\,{\rm d}^3r'\,P_{10}({\bf r})P_{10}({\bf r}')
\left[ g_{20}({\bf r},{\bf r}')\ln g_{20}({\bf r},{\bf r}')
-g_{20}({\bf r},{\bf r}')+1\right]
\nonumber \\
&-& \rho_1\rho_2\int{\rm d}^3r\,{\rm d}^3r'\,P_{10}({\bf r})P_{01}({\bf r}')
\left[ g_{11}({\bf r},{\bf r}')\ln g_{11}({\bf r},{\bf r}')
-g_{11}({\bf r},{\bf r}')+1\right]
\nonumber \\
&-& \frac{1}{2}\rho_2^2
\int{\rm d}^3r\,{\rm d}^3r'\,P_{01}({\bf r})P_{01}({\bf r}')
\left[ g_{02}({\bf r},{\bf r}')\ln g_{02}({\bf r},{\bf r}')
-g_{02}({\bf r},{\bf r}')+1\right]
\nonumber \\
&-& N_1\left( \frac{N_1-1}{2}\ln\frac{N_1}{N_1-1}-\frac{1}{2}\right)
-N_2\left( \frac{N_2-1}{2}\ln\frac{N_2}{N_2-1}-\frac{1}{2}\right) \,.
\label{3-21}
\end{eqnarray}
As a result, i) the contribution from the boundary now grows like
$V\times V^{2/3}N^{-2}\propto N^{-1/3}$; ii) the overall constant
term outside of the integrals is of ${\cal O}(1)$; and iii) the
new integrals now conform to those in the grand-canonical-ensemble
expansion.

For general $(N_1,N_2)$, the constant terms amount to
\begin{eqnarray}
&& \sum_{\shortstack{\scriptsize $n_1+n_2\ge 2$ \\
\scriptsize $n_1\le N_1,\,n_2\le N_2$}}
{N_1\choose n_1}{N_2\choose n_2}
\sum_{\shortstack{\scriptsize $a+b\ge 2$ \\
\scriptsize $a\le n_1,\,b\le n_2$}}
(-1)^{n_1+n_2-a-b}{n_1\choose a}{n_2\choose b}
\nonumber \\
&\times& \ln\left[
\prod_{i_1=0}^{\max\{0,a-1\}}\left( 1-\frac{i_1}{N_1}\right)
\prod_{i_2=0}^{\max\{0,b-1\}}\left( 1-\frac{i_2}{N_2}\right) \right]\,.
\label{3-22}
\end{eqnarray}
For fixed $a$ and $b$ (with $a+b\ge 2$) the prefactor of each
logarithm is the opposite of (\ref{3-16}). Hence, the sum
(\ref{3-22}) equals $\ln(N_1!N_1^{-N_1})+\ln(N_2!N_2^{-N_2})$,
leading to an alternative expression for the entropy MPCE:
\begin{eqnarray}
\frac{S_{N_1,N_2}}{k_B} &=& N_1\left[ \frac{3}{2}-\ln(\rho_1\Lambda_1^3)\right]
+N_2\left[ \frac{3}{2}-\ln(\rho_2\Lambda_2^3)\right]
-N_1\int P_{10}\ln P_{10}-N_2\int P_{01}\ln P_{01}
\nonumber \\
&-& \sum_{\shortstack{\scriptsize $n_1+n_2\ge 2$ \\
\scriptsize $n_1\le N_1,\,n_2\le N_2$}}
{N_1\choose n_1}{N_2\choose n_2}
\sum_{\shortstack{\scriptsize $a+b\ge 2$ \\
\scriptsize $a\le n_1,\,b\le n_2$}}
(-1)^{n_1+n_2-a-b}{n_1\choose a}{n_2\choose b}\int P_{ab}\ln g_{ab}\,,
\nonumber \\
\label{3-23}
\end{eqnarray}
We argue that, using the sum rules (\ref{3-08}), it is always
possible to ``complete'' the integrals $\int P_{ab}\ln g_{ab}$ in
such a way that the integrand behaves like $N^{-a-b}$ at infinity.
In a way analogous to the one-component case, this is accomplished
by adding and subtracting a suitable constant (which is
$-N_1/2-N_2/2$ to second order). If ensemble invariance must hold
for the entropy MPCE, the constant needed is necessarily the
opposite of the number that, added to
\small
\begin{equation}
{N_1\choose n_1}{N_2\choose n_2}
\sum_{\shortstack{\scriptsize $a+b\ge 2$ \\
\scriptsize $a\le n_1,\,b\le n_2$}}
(-1)^{n_1+n_2-a-b}{n_1\choose a}{n_2\choose b}\ln\left[
\prod_{i_1=0}^{\max\{0,a-1\}}\left( 1-\frac{i_1}{N_1}\right)
\prod_{i_2=0}^{\max\{0,b-1\}}\left( 1-\frac{i_2}{N_2}\right) \right] \,,
\label{3-24}
\end{equation}
\normalsize
gives back a ${\cal O}(1)$ quantity as an outcome.

If either $n_1$ or $n_2$ is zero, we are led to the one-component case,
which is treated in appendix A.
Otherwise, when both $n_1$ and $n_2$ are non-zero, the sum in (\ref{3-24})
yields
\small
\begin{eqnarray}
&& \sum_{\shortstack{\scriptsize $a+b\ge 2$ \\
\scriptsize $a\le n_1,\,b\le n_2$}}
(-1)^{n_1+n_2-a-b}{n_1\choose a}{n_2\choose b}\left\{
\sum_{i_1=0}^{\max\{0,a-1\}}\ln\left( 1-\frac{i_1}{N_1}\right) +
\sum_{i_2=0}^{\max\{0,b-1\}}\ln\left( 1-\frac{i_2}{N_2}\right) \right\}
\nonumber \\
&=& \sum_{i_1=0}^{n_1-1}\ln\left( 1-\frac{i_1}{N_1}\right)
\sum_{\shortstack{\scriptsize $a+b\ge 2$ \\
\scriptsize $i_1+1\le a\le n_1,\,b\le n_2$}}
(-1)^{n_1+n_2-a-b}{n_1\choose a}{n_2\choose b}
\nonumber \\
&+& \sum_{i_2=0}^{n_2-1}\ln\left( 1-\frac{i_2}{N_2}\right)
\sum_{\shortstack{\scriptsize $a+b\ge 2$ \\
\scriptsize $a\le n_1,\,i_2+1\le b\le n_2$}}
(-1)^{n_1+n_2-a-b}{n_1\choose a}{n_2\choose b}
\nonumber \\
&=& \underbrace{\sum_{b=0}^{n_2}(-1)^{n_2-b}{n_2\choose b}}_0
\sum_{i_1=1}^{n_1-1}\left[ \ln\left( 1-\frac{i_1}{N_1}\right)
\sum_{a=i_1+1}^{n_1}(-1)^{n_1-a}{n_1\choose a}\right]
\nonumber \\
&+& \underbrace{\sum_{a=0}^{n_1}(-1)^{n_1-a}{n_1\choose a}}_0
\sum_{i_2=1}^{n_2-1}\left[ \ln\left( 1-\frac{i_2}{N_2}\right)
\sum_{b=i_2+1}^{n_2}(-1)^{n_2-b}{n_2\choose b}\right]
\nonumber \\
&=& 0
\label{3-25}
\end{eqnarray}
\normalsize
Hence, in order to produce an ensemble-invariant formula,
a null integral correction must be added to the canonical
correlation integrals associated with the distinct DFs
($n_1=n_2=1$ was such a case, see Eqs.\,(\ref{3-20}) and
(\ref{3-21})).

We reproduce hereafter the entropy MPCE in its final form:
\begin{eqnarray}
\frac{S_{N_1,N_2}}{k_B} &=& N_1\left[ \frac{5}{2}-\ln(\rho_1\Lambda_1^3)\right]
+N_2\left[ \frac{5}{2}-\ln(\rho_2\Lambda_2^3)\right]
-N_1\int P_{10}\ln P_{10}-N_2\int P_{01}\ln P_{01}
\nonumber \\
&-& \sum_{\shortstack{\scriptsize $n_1+n_2\ge 2$ \\
\scriptsize $n_1\le N_1,\,n_2\le N_2$}}
{N_1\choose n_1}{N_2\choose n_2}
\sum_{\shortstack{\scriptsize $a+b\ge 2$ \\
\scriptsize $a\le n_1,\,b\le n_2$}}
(-1)^{n_1+n_2-a-b}{n_1\choose a}{n_2\choose b}\int P_{ab}\ln g_{ab}
\nonumber \\
&-& N_1-N_2\,.
\label{3-26}
\end{eqnarray}
In the r.h.s. of the above formula, the first four terms are the
only ones which survive in the absence of any interaction between
particles and, as such, constitute the ideal contribution to the
entropy. Instead, the last two terms arise from the resummation of
the numbers (\ref{a07}) for each species. When suitably absorbed
into the integrals $\int P_{a0}\ln g_{a0}$ and $\int P_{0b}\ln
g_{0b}$, these terms eventually make the canonical-ensemble
entropy expansion look like the grand-canonical one. In
particular, for a large ($N_1,N_2\gg 1$) and {\em homogeneous}
binary system, the entropy expansion starts as follows:
\begin{eqnarray}
\frac{S_{N_1,N_2}}{k_B} &=& N_1\left[ \frac{5}{2}-\ln(\rho_1\Lambda_1^3)\right]
+N_2\left[ \frac{5}{2}-\ln(\rho_2\Lambda_2^3)\right]
\nonumber \\
&-& \frac{1}{2}\rho_1^2
\int{\rm d}^3r\,{\rm d}^3r'
\left[ g_{20}({\bf r},{\bf r}')\ln g_{20}({\bf r},{\bf r}')
-g_{20}({\bf r},{\bf r}')+1\right]
\nonumber \\
&-& \rho_1\rho_2\int{\rm d}^3r\,{\rm d}^3r'
\left[ g_{11}({\bf r},{\bf r}')\ln g_{11}({\bf r},{\bf r}')
-g_{11}({\bf r},{\bf r}')+1\right]
\nonumber \\
&-& \frac{1}{2}\rho_2^2
\int{\rm d}^3r\,{\rm d}^3r'
\left[ g_{02}({\bf r},{\bf r}')\ln g_{02}({\bf r},{\bf r}')
-g_{02}({\bf r},{\bf r}')+1\right] +\ldots
\label{3-27}
\end{eqnarray}
If the above expansion is truncated after its pair-correlation
terms, this formula provides a low-density approximation for the
entropy of the mixture.

The MPCE of the entropy derived above applies to a binary mixture
of interacting {\em point} particles. However, the formula does
not change even if the particles possess further degrees of
freedom. In fact, the same comments that we made at the end of
section 2 still apply. A case of this sort is a mixture of spheres
and spherocylinders. We shall analyze in a future publication the
phase diagram of this model system in terms of the RMPE. Here, we
focus our attention on the general structure of the {\em distinct}
two-body reduced DF, since the case of the sphere-sphere DF is
obvious while the symmetries of the self DF for spherocylinders
have been already discussed in section 2.

In order to find the maximum number of independent scalar
variables that intervene in the calculation of $g_{11}(\xi,{\bf q})$,
it is convenient to work in a reference system where the
spherocylinder (species 1) is centred at the origin and lies
along the $z$ axis. In this reference frame, the position of the
sphere (species 2) can be parameterized in terms of two variables
only, namely the length $r_{12}$ and colatitude $\vartheta_{12}$
of the vector joining the two centres of mass (as usual,
$\varphi_{12}=0$ for a convenient choice of the $x$ axis). The
same result is obviously obtained when viewing the situation from
a reference system where the sphere is centred at the origin. In
this case, the need for two further $\theta_{12}$ and $\phi_{12}$
variables (see section 2) is only apparent, since the $z$ and the
$x$ axis can be chosen in such a way that such two variables
vanish altogether. If the axes of the spherocylinders are frozen
and parallel to each other, no further simplification occurs, and
the distinct two-body DF is again a function of $r_{12}$ and
$\vartheta_{12}$.

The numerical calculation of $g_{11}$ for a homogeneous mixture is
carried out as follows. After taking $\xi=({\bf r},\theta,\phi)$,
${\bf x}={\bf q}-{\bf r}$, and invoking homogeneity, we have:
\begin{eqnarray}
\frac{\rho_1}{4\pi}\rho_2\,g_{11}(\xi,{\bf q}) &=&
\frac{1}{4\pi V}\int{\rm d}^5\xi\left< \sum_{i=1}^{N_1}\sum_{j=1}^{N_2}\,
\delta^5(\Xi_i-\xi)\delta^3({\bf Q}_j-{\bf q})\right>
\nonumber \\
&=& \frac{1}{4\pi V}\left< \sum_{i,j}\int{\rm d}\theta\,{\rm d}\phi\,
\delta(\Theta_i-\theta)\delta(\Phi_i-\phi)
\int{\rm d}^3r\,\delta^3({\bf R}_i-{\bf r})
\delta^3({\bf Q}_j-{\bf r}-{\bf x})\right>
\nonumber \\
&=& \frac{N_1}{4\pi V}\left< \sum_{j=1}^{N_2}
\delta^3({\bf Q}_j-({\bf R}_1+{\bf x}))\right> \,.
\label{3-28}
\end{eqnarray}
Hence, $\rho_2\,g_{11}(\xi,{\bf q})$ is the average {\em density}
of spheres in ${\bf R}_1+{\bf x}$ when a spherocylinder is
centred in ${\bf R}_1$. An explicit formula is:
\begin{equation}
g_{11}(r_{12},\vartheta_{12})\simeq\frac{\Delta{\cal
N}_2(r_{12},\vartheta_{12})} {\rho_2\cdot
2\pi r_{12}^2\sin\vartheta_{12}\Delta r_{12}\Delta\vartheta_{12}}\,,
\label{3-29}
\end{equation}
$\Delta{\cal N}_2(r_{12},\vartheta_{12})$ being the number of
spheres within a tiny spherical ring of volume
$2\pi r_{12}^2\sin\vartheta_{12}\Delta r_{12}\Delta\vartheta_{12}$,
centred at the position specified by $r_{12}$ and $\vartheta_{12}$.

\newpage
%
%
\section{Conclusions}

In this paper, we have outlined a constructive method for building
up the entropy multiparticle-correlation expansion in the
canonical ensemble, term by term, for both pure and mixed systems
of classical particles. The aim of this effort is twofold: i) to
unveil the hidden combinatorial structure behind the expansion;
ii) to set the stage for an application of the entropy-based
ordering criterion introduced by Giaquinta and coworkers to the
phase diagram of a binary mixture of hard spheres and
spherocylinders. In this respect, we have discussed here the
general symmetries owned by the three pair distribution functions.
A detailed analysis of the phase diagram of the model in the
framework provided by the zero-RMPE criterion will be the object
of a forthcoming publication.

\newpage
\appendix
\section{Ensemble invariance of the entropy MPCE}
\setcounter{equation}{0}
\renewcommand{\theequation}{A.\arabic{equation}}

In this appendix, an argument appearing in Ref.\,\cite{Prestipino}
is reproduced for the reader's convenience. This argument deals
with the overall constant term that appears in Eq.\,(\ref{2-12})
when we eliminate the normalized DFs in favour of the reduced DFs
through Eq.\,(\ref{2-06}).

For one-component systems, this constant amounts to:
\begin{equation}
\sum_{n=2}^N{N\choose n}\sum_{a=2}^n(-1)^{n-a}{n\choose a}\ln\frac{(N-1)
(N-2)\cdots(N-a+1)}{N^{a-1}}\,.
\label{a01}
\end{equation}
For each $a$ value, the prefactor of the respective logarithm is just the
number (\ref{2-14}).
Hence, the sum (\ref{a01}) equals $\ln(N!N^{-N})$, thus yielding a new
form of the entropy MPCE:
\begin{equation}
\frac{S_N}{k_B}=N\left[ \frac{3}{2}-\ln(\rho\Lambda^3)\right]
-N\int P_1\ln P_1-\sum_{n=2}^N{N\choose n}\sum_{a=2}^n(-1)^{n-a}
{n\choose a}\int P_{1\ldots a}\ln g_{1\ldots a}\,,
\label{a02}
\end{equation}
where $\rho=N/V$.

We now show that any single term in the sum over $n$ at (\ref{a01}) is
extensive.
In fact, we have first:
\begin{equation}
{N\choose n}\sum_{a=2}^n(-1)^{n-a}{n\choose a}\sum_{k=1}^{a-1}
\ln\left( 1-\frac{k}{N}\right) =
{N\choose n}\sum_{k=1}^{n-1}\left[ \ln\left( 1-\frac{k}{N}\right)
\sum_{a=k+1}^n(-1)^{n-a}{n\choose a}\right] \,.
\label{a03}
\end{equation}
We prove in appendix B that:
\begin{equation}
\sum_{a=k+1}^n(-1)^{n-a}{n\choose a}=(-1)^{n-1-k}{n-1\choose k}\,.
\label{a04}
\end{equation}
Then,
\begin{eqnarray}
&& {N\choose n}\sum_{a=2}^n(-1)^{n-a}{n\choose a}\ln\frac{(N-1)
(N-2)\cdots(N-a+1)}{N^{a-1}}
\nonumber \\
&=& -{N\choose n}\sum_{k=1}^{n-1}(-1)^{n-k}{n-1\choose k}
\ln\left( 1-\frac{k}{N}\right)
\nonumber \\
&=& {N\choose n}\left\{ \frac{1}{N}\sum_{k=1}^{n-1}(-1)^{n-k}{n-1\choose k}k+
\frac{1}{2N^2}\sum_{k=1}^{n-1}(-1)^{n-k}{n-1\choose k}k^2+\ldots\right\} \,.
\nonumber \\
\label{a05}
\end{eqnarray}
In appendix B, we also show that
\begin{equation}
\sum_{k=1}^{n-1}(-1)^{n-k}{n-1\choose k}k^d=\left\{
\begin{array}{rl}
0\,, & \,\,\,{\rm for}\,\,d=1,\ldots,n-2 \\
-(n-1)!\,, & \,\,\,{\rm for}\,\,d=n-1
\end{array}
\right.\,.
\label{a06}
\end{equation}
In conclusion, we obtain:
\begin{equation}
{N\choose n}\sum_{a=2}^n(-1)^{n-a}{n\choose a}\ln\frac{(N-1)
(N-2)\cdots(N-a+1)}{N^{a-1}}\sim -\frac{N}{n(n-1)}\,.
\label{a07}
\end{equation}

As discussed in the main text, the integrals in Eq.\,(\ref{a02})
cannot be easily computed numerically since, for finite $N$, the
system boundary also contributes in a significant way. However,
upon taking advantage of the canonical-ensemble sum rules for the
reduced DFs (see Eqs.\,(\ref{2-07})), it should be always possible
to make every integrand in Eq.\,(\ref{a02}) sufficiently small at
large distances. This is accomplished by adding (and subtracting)
a quantity equal to the number in (\ref{a07}), with the result of
leaving an overall ${\cal O}(1)$ number outside of the integral.
Furthermore, the new form of the integral can be made identical to
the (so-called) fluctuation integral of the same order which
appears in the grand-canonical-ensemble expansion.

We show this explicitly for the third-order term in the entropy expansion,
which, when expressed in terms of the reduced DFs, reads:
\begin{equation}
-{N\choose 3}\ln\frac{(N-1)^2}{N(N-2)}-\frac{1}{6}\rho^3
\int{\rm d}^3r_1\,{\rm d}^3r_2\,{\rm d}^3r_3\,P_1P_2P_3\,
g_{123}\ln\frac{g_{123}}{g_{12}g_{13}g_{23}}\,.
\label{a08}
\end{equation}
In order to conform to the grand-canonical-ensemble expansion, we
have to add the integral
\begin{equation}
-\frac{1}{6}\rho^3
\int{\rm d}^3r_1\,{\rm d}^3r_2\,{\rm d}^3r_3\,P_1P_2P_3\,
(-g_{123}+3g_{12}g_{13}-3g_{12}+1)
\label{a09}
\end{equation}
which, in view of the canonical-ensemble sum rules (\ref{2-04})
and (\ref{2-07}), is equal to $-N/6$. This number is exactly the
same constant that must be subtracted to
\begin{equation}
-{N\choose 3}\ln\frac{(N-1)^2}{N(N-2)}
\label{a10}
\end{equation}
in order to produce an ${\cal O}(1)$ constant. As a {\it caveat},
we note that
\begin{equation}
-\frac{1}{6}\rho^3
\int{\rm d}^3r_1\,{\rm d}^3r_2\,{\rm d}^3r_3\,P_1P_2P_3\,
\left[ \kappa g_{123}+(1-2\kappa)g_{12}g_{13}+(\kappa-2)g_{12}+1\right] =
-\frac{N}{6}
\label{a11}
\end{equation}
for any real $\kappa$, not simply $-1$.

\newpage
\section{Two combinatorial identities}
\setcounter{equation}{0}
\renewcommand{\theequation}{B.\arabic{equation}}

In this appendix, the formulae (\ref{a04}) and (\ref{a06}) are
proved by induction.

First, we prove that, for any $n\ge 2$ and $1\le k\le n-1$:
\begin{equation}
\sum_{a=k+1}^n(-1)^{n-a}{n\choose a}=(-1)^{(n-1)-k}{n-1\choose k}\,.
\label{b01}
\end{equation}
Equation (\ref{b01}) is valid also for $k=0$ and any $n\ge 1$. We
argue by induction over $n$. For $n=2$ and $k=1$, the formula
(\ref{b01}) is trivially correct. Then, assuming that the formula
is correct for an arbitrary fixed $n$ and all positive $k<n$, we
calculate the l.h.s. of Eq.\,(\ref{b01}) for $n+1$ and any $1\le
k\le n-1$ (the case $k=n$ is obvious):
\begin{eqnarray}
&& \sum_{a=k+1}^{n+1}(-1)^{(n+1)-a}{n+1\choose a}=
-\sum_{a=k+1}^n(-1)^{n-a}\left[ {n\choose a}+{n\choose a-1}\right] +1
\nonumber \\
&=& -(-1)^{(n-1)-k}{n-1\choose k}+\sum_{a=k}^n(-1)^{n-a}{n\choose a}
=(-1)^{n-k}\left[ {n-1\choose k}+{n-1\choose k-1}\right]
\nonumber \\
&=& (-1)^{n-k}{n\choose k}\,.
\label{b02}
\end{eqnarray}
which is just the r.h.s. of Eq.\,(\ref{b01}), but for $n+1$ which
replaces $n$.

Next, we show that, for $n\ge 3$:
\begin{equation}
\sum_{k=1}^{n-1}(-1)^{n-k}{n-1\choose k}k^d=\left\{
\begin{array}{rl}
0\,, & \,\,\,{\rm for}\,\,d=1,\ldots,n-2 \\
-(n-1)!\,, & \,\,\,{\rm for}\,\,d=n-1
\end{array}
\right.\,,
\label{b03}
\end{equation}
while it is trivial to check that
\begin{equation}
\sum_{k=1}^{n-1}(-1)^{n-k}{n-1\choose k}k^d=-1\,\,\,\,\,\,
(n=2,d=1)\,.
\label{b04}
\end{equation}
Arguing inductively, let us suppose that Eq.\,(\ref{b03}) is valid
for a given value $n$ (checking this for $n=3$ is immediate), and
see what happens for $n+1$:
\begin{eqnarray}
&& \sum_{k=1}^n(-1)^{n+1-k}{n\choose k}k^d=
-\sum_{k=1}^{n-1}(-1)^{n-k}{n\choose k}k^d-n^d
\nonumber \\
&=& -\sum_{k=1}^{n-1}(-1)^{n-k}\left[ {n-1\choose k}+
{n-1\choose k-1}\right] k^d-n^d
\nonumber \\
&=& -\sum_{k=1}^{n-1}(-1)^{n-k}{n-1\choose k}k^d+
\sum_{k=0}^{n-2}(-1)^{n-k}{n-1\choose k}(k+1)^d-n^d
\nonumber \\
&=& -\sum_{k=1}^{n-1}(-1)^{n-k}{n-1\choose k}k^d+
\sum_{k=1}^{n-1}(-1)^{n-k}{n-1\choose k}(k+1)^d+(-1)^n
\nonumber \\
&=& -\sum_{k=1}^{n-1}(-1)^{n-k}{n-1\choose k}k^d+
\sum_{k=1}^{n-1}(-1)^{n-k}{n-1\choose k}\sum_{m=0}^d{d\choose m}k^m+(-1)^n
\nonumber \\
&=& -\sum_{k=1}^{n-1}(-1)^{n-k}{n-1\choose k}k^d+\sum_{m=0}^d{d\choose m}
\left\{ \sum_{k=1}^{n-1}(-1)^{n-k}{n-1\choose k}k^m\right\} +(-1)^n
\nonumber \\
\label{b05}
\end{eqnarray}
Focussing on this intermediate result, we distinguish three cases:

i)\,\, For $1\le d<n-1$, the first term is zero as is the second for $m>0$.
As for the rest:
\begin{eqnarray}
&& \sum_{k=1}^{n-1}(-1)^{n-k}{n-1\choose k}+(-1)^n=
-\sum_{k=1}^{n-1}(-1)^{(n-1)-k}{n-1\choose k}+(-1)^n
\nonumber \\
&=& -\sum_{k=0}^{n-1}(-1)^{(n-1)-k}{n-1\choose k}=-(1-1)^{n-1}=0\,.
\label{b06}
\end{eqnarray}

ii)\,\, For $d=n-1$ instead, the first term is $(n-1)!$, while the second is
zero for $0<m<d$. The rest equals to:
\begin{eqnarray}
&& (n-1)!+\sum_{k=1}^{n-1}(-1)^{n-k}{n-1\choose k}k^{n-1}+
\sum_{k=1}^{n-1}(-1)^{n-k}{n-1\choose k}+(-1)^n
\nonumber \\
&=& \sum_{k=1}^{n-1}(-1)^{n-k}{n-1\choose k}+(-1)^n=0\,,
\label{b07}
\end{eqnarray}
like in the previous case.

iii)\,\, Finally, and given all the above results, for $d=n$ we have:
\begin{eqnarray}
&& -\sum_{k=1}^{n-1}(-1)^{n-k}{n-1\choose k}k^n+
{n\choose 0}\sum_{k=1}^{n-1}(-1)^{n-k}{n-1\choose k}+
\sum_{m=1}^{n-2}{n\choose m}\sum_{k=1}^{n-1}(-1)^{n-k}{n-1\choose k}k^m
\nonumber \\
&-& {n\choose n-1}(n-1)!+
{n\choose n}\sum_{k=1}^{n-1}(-1)^{n-k}{n-1\choose k}k^n+(-1)^n=-n!\,,
\label{b08}
\end{eqnarray}
as we wanted to show.

\newpage
%
%

\newpage
%
%
\begin{center}
\large
FIGURE CAPTION
\normalsize
\end{center}
\begin{description}
\item[{\bf Fig.\,1 :}] The Euler angles $\theta,\phi$, and $\psi$:
The axes $x,y$, and $z$ form the laboratory reference frame,
whereas $x',y'$, and $z'$ are parallel to the body set of axes.
$\zeta$, called nodal line, is the straight line perpendicular to
$z$ and $z'$. It can also be viewed as the intersection between
the $xy$ and $x'y'$ planes. The $z'$ axis is chosen so as to
coincide with the symmetry axis of the molecule (represented in
the picture as a rod). The Euler angles are in the ranges
$0\le\theta<\pi$, $0\le\phi<2\pi$, and $0\le\psi<2\pi$.
\end{description}
\newpage
%
%
\begin{figure}
\begin{center}
\setlength{\unitlength}{1cm}
\begin{picture}(18,15)(0,0)
\put(-1.5,0){\psfig{file=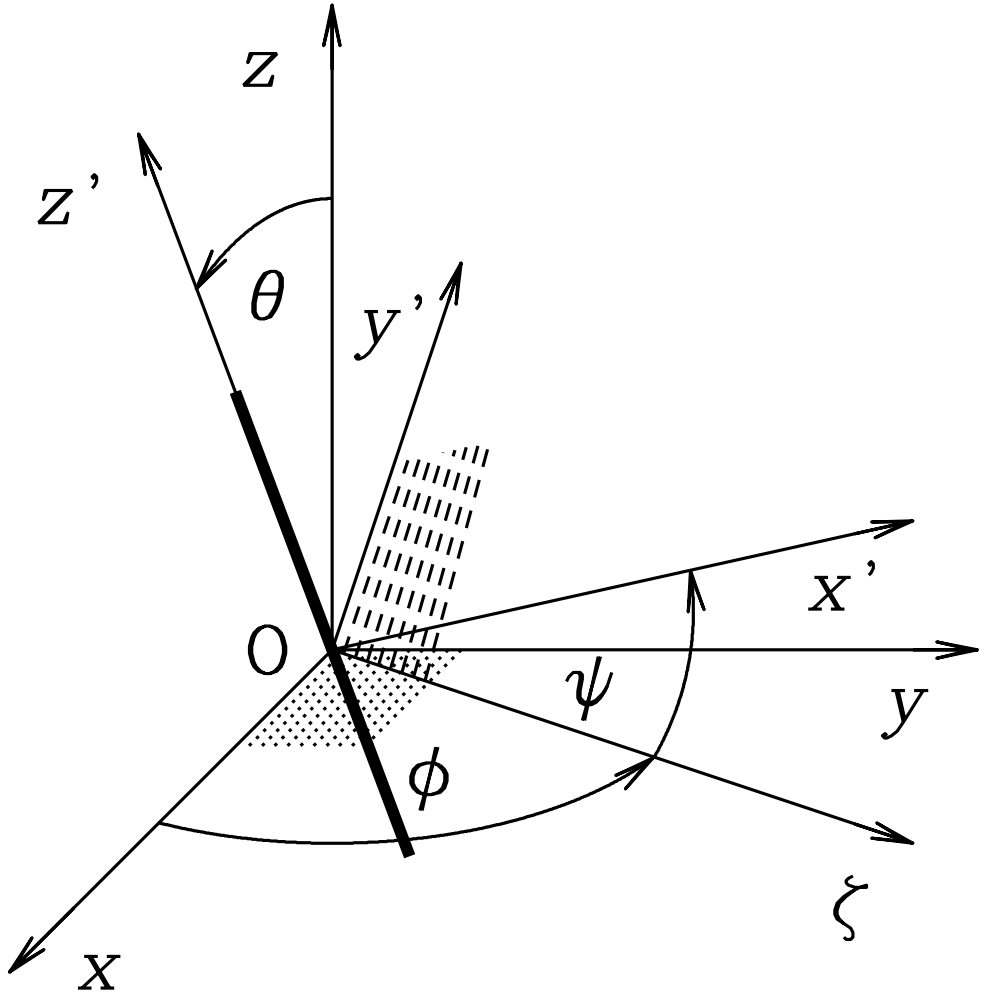,width=18cm,bbllx=0cm}}
\end{picture}
\caption[1]{
The Euler angles $\theta,\phi$, and $\psi$:
The axes $x,y$, and $z$ form the laboratory reference frame,
whereas $x',y'$, and $z'$ are parallel to the body set of axes.
$\zeta$, called nodal line, is the straight line perpendicular to
$z$ and $z'$. It can also be viewed as the intersection between
the $xy$ and $x'y'$ planes. The $z'$ axis is chosen so as to
coincide with the symmetry axis of the molecule (represented in
the picture as a rod). The Euler angles are in the ranges
$0\le\theta<\pi$, $0\le\phi<2\pi$, and $0\le\psi<2\pi$.
}
\end{center}
\end{figure}

\end{document}